\documentclass[12pt]{article}

\def\pa{\partial}
\def\d{\delta}
\def\tr{\mbox{tr}}

\def\be{\begin{equation}}
\def\ee{\end{equation}}
\newcommand{\bea}{\begin{eqnarray}}
\newcommand{\eea}{\end{eqnarray}}
\begin {document}
\begin{titlepage}
January 2002 \\
\begin{flushright}
HU Berlin-EP-02/02\\
\end{flushright}
\mbox{ }  \hfill hep-th/0201161
\vspace{5ex}
\Large
\begin {center}
{\bf
Gauge invariant operators in field theories on non-commutative spaces}
\footnote{To appear in the Proceedings of the RTN meeting ``The Quantum Structure of Spacetime and the Geometric Nature of Fundamental Interactions'',
Corfu, September 13-20, 2001}
\end {center}
\large
\vspace{1ex}
\begin{center}
Harald Dorn \footnote{dorn@physik.hu-berlin.de}
\end{center}
\begin{center}
Humboldt--Universit\"at zu Berlin, Institut f\"ur Physik\\
Invalidenstra\ss e 110, D-10115 Berlin\\[2mm]
\end{center}
\vspace{4ex}
\rm
\begin{center}
{\bf Abstract}
\end{center}
We review some selected aspects of the construction of gauge invariant operators in field theories on non-commutative spaces and their relation to the energy
momentum tensor as well as to the non-commutative loop equations.
\vfill
\end{titlepage}
\section{Introduction}
We consider $U(N)$ gauge theory on the simplest non-commutative
space, flat space with the commutation relations
\be
[x^{\mu},x^{\nu}]~=~i\theta ^{\mu\nu}~
\label{1}
\ee
among its coordinates. Throughout the following discussions use is made of
the $\star $-product formulation. The gauge field transforms under gauge
transformations as
\be
A_{\mu}~\rightarrow ~\Omega\star A_{\mu}\star \Omega ^{\dagger}~+~
i~\Omega\star\partial _{\mu}\Omega ^{\dagger}~,~~~~~~\Omega\star\Omega ^{\dagger}=1~
\label{2}
\ee
and the field strength $F_{\mu\nu}=\partial _{\mu}A_{\nu}-\partial _{\nu}A_{\mu}-iA_{\mu}\star A_{\nu}+iA_{\nu}\star A_{\mu}$ as
\be
F_{\mu\nu}~\rightarrow \Omega\star F_{\mu\nu}\star\Omega ^{\dagger}~.
\label{3}
\ee
The non-commutative $\star $-product under integration over the whole space
obeys
\be
\int dx ~f\star g~=~\int dx ~fg~=~\int dx ~ g\star f~.
\label{4}
\ee
Local operators like $\tr F^2$
being gauge invariant in standard gauge
theories are no longer gauge invariant, however due to (\ref{4})
gauge invariance is restored after integration over the whole space.

The situation is similarly for the Wilson functionals
\be
U [C,A] = P_\star \exp \left( i \int_0^1 A_\mu (\xi (t))\frac{d\xi ^{\mu}}
{dt }dt
 \right)
\label{5}
\ee
where
$P_\star$ denotes path ordering along the contour $C$ defined by $\xi (t )$ from right to left with respect to increasing $t $ of $\star$-products of functions. The star multiplication is performed with respect to a constant mode in
$\xi (t )$.\footnote{If $\xi (t )=x+\eta (t )=x'+\eta '(t)$
taking the $\star $-products w.r.t. $x$ or $x'$ yields the same result.}
Under gauge transformations $U[C,A]$ transforms as
\be
U[C,A]~\rightarrow ~ \Omega (\xi (1))\star U[C,A]\star \Omega ^{\dagger}(\xi (0))~.
\label{6}
\ee
For closed contours $\xi (0)=\xi (1)$ besides the trace one has again to perform an integration over the whole space to get a gauge invariant quantity.

In addition there appears the new possibility to relate gauge invariant objects
also to open contours \cite{iikk}
\be
W[C,A]~=~\frac{1}{N}\int dx~ \tr ~U[x+C,A]\star e^{-ik_{\xi}x}~~~\mbox{with}~~
\xi (1)-\xi (0)=\theta k_{\xi}~.
\label{7}
\ee
The gauge invariance of $W$ is a consequence of (\ref{4}),(\ref{6}) and
\be
e^{ikx}\star f(x)\star e^{-ikx}~=~f(x+\theta k)~.
\label{8}
\ee
\section{Localized gauge invariant operators}
Here one starts with local
\footnote{Here we understand local on a technical level as
operators built out of $\star $ powers of the gauge field and its derivatives.}
${\cal O}(x)$ transforming under (\ref{2})
in the adjoint, i.e. ${\cal O}(x)\rightarrow\Omega (x)\star{\cal O}(x)\star
\Omega ^{\dagger}$. Then the trace of the Fourier transform
$$
\int dx~\tr{\cal O}(x)e^{-ikx}
$$
is gauge invariant only for $k=0$. To construct some generalization which is
invariant for all $k$ we follow \cite{dr,ghi} and repair the mismatch of
the gauge functions $\Omega $ by inserting a suitable adapted Wilson
functional, i.e.
\be
\tr\widetilde{\cal O}(k)~=~ \int dx~\tr (U[x+C,A]\star{\cal O}(x))
\star e^{-ikx}~.
\label{9}
\ee
The above construction is gauge invariant for each contour $C$ with
\be
\xi (1)~=~\theta k~,~~~~~\xi(0)~=~0.
\label{10}
\ee
Applying to $\tr\widetilde{\cal O}(k)$ the usual inverse Fourier
transformation one arrives at the gauge invariant coordinate space operator
\be
\tr\widehat{\cal O}(y)~=~ \frac{1}{(2\pi )^D}\int dk~\tr\widetilde{\cal O}(k)~e^{iky}
~.
\label{11}
\ee
Among the contours $C$ obeying (\ref{10}) the straight ones are distinguished.
Only then the construction with ${\cal O}$ inserted at an endpoint can
equally be replaced by a setup where ${\cal O}$ is inserted at an arbitrary
point of the contour \cite{ghi}. Another benefit of using straight contours
in (\ref{9}) is related to the use of covariant coordinates in the sense
of ref.\cite{mssw}
\be
X^{\mu}~=~x^{\mu}~+~(\theta A(x))^{\mu}~.
\label{12}
\ee
One can prove for the exponential $\star$-power series of $-ikX$ the remarkable identity \cite{dw}
\be
e_{\star}^{-ikX}~=~e^{-ikx}\star U(k,x)~,
\label{13}
\ee
with
\be
U(k,x)~=~U[x,C]~,~~~~C:~\xi(t)~=~\theta kt~.
\label{13a}
\ee
Then the construction of $\widehat{\cal O}(x)$ out of ${\cal O}(x)$
can be summarized by
\be
\tr\widehat{\cal O}(y)~=~\int dx~\tr{\cal O}(x)~\delta _{\star} (X-y)~.
\label{14}
\ee
Replacing the $\delta _{\star}$-function by some smooth regularization one
gets the pseudo localized operators studied in more detail in ref.\cite{bl}.

\section{Energy momentum tensor}
The energy momentum tensor of non-commutative gauge theories has been studied
both from the string theoretical \cite{oo} and the field theoretical \cite{ad}
point of view. The resulting expressions are different since the leading
order in $\alpha '$ studied so far is not seen on the pure field theoretical
level. Therefore it would be interesting to extend the string calculation
to the next-leading contributions. We now comment the field theory analysis.

Applying the Noether procedure combined with a suitably adapted gauge
transformation one gets
\be
D_{\mu}~T^{\mu\nu}~=~\partial _{\mu}T^{\mu\nu}~-~iA_{\mu}\star T^{\mu\nu}
~+~iT^{\mu\nu}\star A_{\mu}~=0~,
\label{15}
\ee
with
\be
T^{\mu\nu}~=~2\{F^{\mu\rho},F^{\nu}_{\ \rho}\}_\star -\eta ^{\mu\nu}
F^{\alpha\beta}\star F_{\alpha\beta}~.
\label{16}
\ee
This symmetric tensor $T^{\mu\nu}$ even after taking the trace is not gauge invariant. It also does not
fulfill the standard local conservation law.
With the technique presented in the last chapter we find the gauge invariant
tensor
\be
\hat T^{\mu\nu}(y)~=~\frac{1}{(2\pi )^D}\int dkdx~e^{iky}
e^{-ikx}\star \tr\big ( U(k,x)\star T^{\mu\nu}(x)\big )~.
\label{17}
\ee
$\partial _{\mu}\hat T^{\mu\nu}(y)$ turns out to be different from zero
but equal to a derivative. Therefore after a straightforward redefinition
we end up with
\bea
{\bf T}^{\mu\nu}(y)&= &\frac{1}{(2\pi )^D}\int dkdx~e^{iky}
e^{-ikx}\star \tr\Big [ U(k,x)\star T^{\mu\nu}(x)\nonumber\\
&&-~\theta ^{\mu\alpha}P_{\star}\left (
\int _0^1ds F_{\alpha\beta}(x+s\theta k)U(k,x)T^{\beta\nu}(x)\right )\Big ]~ .
\label{18}
\eea
as our energy momentum tensor. ${\bf T}^{\mu\nu}$ is gauge invariant and
locally conserved
\be
\partial _{\mu}{\bf T}^{\mu\nu}~=~0 ,
\label{19}
\ee
The price for enforcing (\ref{19}) is the loss of the symmetry of the
tensor. For more discussions of the interplay between local conservation
and symmetry see \cite{ad}.
\section{Loop equations}
In the previous sections we concentrated ourselves on the
use of Wilson functionals as building blocks in the construction
of localized gauge invariant operators. Now we look closer on the dependence
of the Wilson functionals on the shape of the contours. In analogy to
standard Yang-Mills gauge theories one expects that the dynamics
can be encoded in equations containing second variational derivatives
with respect to the contour. We now closely follow the steps known
in the standard case \cite{loop}. The geometrical setting to derive
a formula for the derivative with respect to the insertion of an area
derivative is unchanged, hence
\be
\partial ^\mu \frac{\d W [C] }{\d \sigma ^{\mu\nu} (\xi (t ))}= \frac{i}{N}
\int dx \tr P_\star
\left( D^\mu F_{\mu\nu} (x+\xi (t ) )
\exp \left( i\int_C A_\mu (x+\xi (s ))d\xi^\mu {(s)}
 \right) \right).
\label{20}
\ee
In the quantized version of the standard Yang-Mills theory the equation of
motion gives rise to contact terms constituting a nontrivial r.h.s.
of the loop equations \cite{loop}. Under the vacuum expectation value
translation invariance is restored as usual. This implies
a trivial divergence for the vev of (\ref{7}) which we cancel by dividing
out the space-time volume $V$. Furthermore, a careful analysis of the
modifications
introduced by the $\star $-product leads for the vev of $W$ for a closed
contour\footnote{Below we use a subscript $c$ and $o$ for emphasizing the
closed and open nature of the contour.} to \cite{ad}
\be
\frac{1}{V} \pa^\mu \frac{\d}{\d \sigma ^{\mu\nu} (\xi )} \langle W_c[C]
\rangle
 = - \frac{g^2N}{V} \frac{1}{(2\pi )^D \det \theta } \int_C d\eta_\nu
\langle W_o [C_{\xi \eta}] W_o [C_{\eta \xi }] \rangle .
\label{21}
\ee
$C_{\eta \xi}$ denotes the part of $C$ between $\xi $ and $\eta $.
Relative to the commutative case where on the r.h.s. only
$\xi =\eta $ contributes, now the former contact terms become
smeared in some sense, and we have contributions from all
$\eta $.
Separating connected and disconnected parts for the correlator on the
r.h.s. one finally gets
\bea
\frac{1}{V}\partial ^{\mu}\frac{\delta }{\delta\sigma ^{\mu\nu}(\xi )}
\langle W_c[C]\rangle & = & -\frac{g^2N}{ V^2}\int _C d\eta_{\nu}~
\delta (\xi -\eta )~\langle W_c[C_{\xi \eta }]\rangle~\langle
W_c[C_{\eta \xi}]\rangle \nonumber \\ & & -\frac{g^2N}{(2\pi )^D
V\det\theta}~
\int _C d\eta _{\nu}~\langle W_o[C_{\xi\eta}]~W_o[C_{\eta\xi}]
\rangle_{conn}~.~~~
\label{22}
\eea
It is remarkable
that for finite $N$ the new gauge invariant objects for open contours
appear to be necessary for the description of the dynamics of closed
loops. In the t'Hooft limit ($N\rightarrow\infty ,~~g^2N$ fix) the
second term on the r.h.s. is suppressed by a relative $1/N^2$ factor with
respect to the first one resulting in just the same equation as in the commutative case.

For the two-point function of $W$'s for closed contours there appears of
course on the
r.h.s. one term in which the second contour has a pure spectator role. Due
to the smearing of the contact term there contributes still another term,
even if the two contours have no point in common \cite{ad}
\bea
\frac{1}{V}\partial ^{\mu}\frac{\delta }{\delta\sigma ^{\mu\nu}(\xi ^1)}
\langle W_c[C^1]W_c[C^2]\rangle =~~~~~~~~~~~~~~~~~~~~~~~~~~~~~~~~~~~~~~~~~~~~~~~~~~~~ \label{23}\\
-\frac{g^2N}{(2\pi )^{D}V\det\theta}~
\int _{C^1}d\eta ^1_{\nu}~\langle W_o[C^1_{\xi ^1\eta ^1}]~W_o[C^1_{\eta ^1\xi ^1}]~W_c[C^2]\rangle
\nonumber\\
-\frac{g^2}{NV}\int _{C^2}d\eta ^2_{\nu}~\langle ~W_c[C^1_{\xi ^1}\circ (C^2_{\eta ^2}+\xi ^1 -\eta ^2 )]~\rangle ~.
\nonumber
\eea
Above $C^1_{\xi}\circ (C^2_{\eta}+\xi -\eta )$ denotes the closed contour
obtained by starting at $\xi $ on $C^1$ going along the whole $C^1$ back to
$\xi $ and then along the with $(\xi -\eta )$ shifted version of $C^2$.\\

The one-point function of $W$ for open contours vanishes since
translation invariance leads to a factor $\delta (k)$. The correlators
of $W$'s for several open contours are nonvanishing if the related
momenta sum up to zero. These correlation functions have
been studied in ref. \cite{dk}. Let us first introduce some notation.
By $C_{(s )}= \{\xi_{s}(t ),~0\leq t\leq 1\}$ we denote the following contour tailored out of $C=\{\xi (t ),~0\leq t\leq 1\}$
\bea
\xi _{s}(t )&=&\xi (s +t )-\xi (1)+\xi (0),~~0\leq t\leq 1-s \nonumber\\
\xi _{s}(t )&=&\xi (t -1+s ),~~1-s \leq t\leq 1~.
\label{24}
\eea
If $C$ is closed $C_{(s )}$ is equal to $C_{\xi (s )}$ in the sense
used in (\ref{23}).
In addition we use the notion $C_{(st)}$ for the part of $C$ between
$\xi (\mbox{min}(s ,t ))$ and $\xi (\mbox{max}(s ,t ))$ as well
as $C/C_{(st )}$ for the contour obtained from  $C$ by cutting out
$C_{(st )}$ and gluing together the remaining two parts after a suitable translation of one of the partners. $(k_1)_{(st )}$ stands for
the momentum related to $C^1_{(st )}$. Then the equation for the two-point function, after making
use of the cyclic symmetry \cite{dk}
\be
W[C]~=~W[C_{(s )}]~,
\label{25}
\ee
takes the form
\bea
\frac{1}{V}\partial ^{\mu}\frac{\delta }{\delta \sigma ^{\mu\nu}(\xi ^1(t ))}
\langle W_o[C^1]W_o[C^2]\rangle =~~~~~~~~~~~~~~~~~~~~~~~~~~~~~~~~~~~~~~~~~~~~~
~~~~~~~ \label{26}\\
-\frac{g^2N}{(2\pi )^{D}V\det\theta}~
\int _{C^1}d\xi ^1_{\nu}(s )~\langle W_o[C^1_{(st )}]~W_o[C^1/
C^1_{st }]~W_o[C^2]\rangle ~e^{-\frac{i}{2}k_1\theta (k_1)_{(st )}}
\nonumber\\
-\frac{g^2}{NV}\int _{C^2}d\xi ^2_{\nu}(s )~\langle ~W_c[C^1_{(t )}\circ (C^2_{(s )}+\xi ^1(t ) -\xi ^2 (s )-\xi ^1(1)+\xi ^1(0))]~\rangle ~ e^{-ik_2(\xi ^1(t )-\xi ^2(s ))}~ .
\nonumber
\eea
In the limit of closed contours $C^1$ and $C^2$ (\ref{26}) equals
(\ref{23}).\footnote{After adapting the notation: $\xi ^1(t)$ is called
$\xi ^1$ and $\xi ^i(s)$ $\eta ^i$ in (\ref{23}).}
Note further that $k_1+k_2=0$ always implies $\xi ^1(1)-\xi ^1(0)=\xi ^2(0)-\xi ^2(1)$. Therefore the second term on the r.h.s. involves the Wilson functional for a
closed contour.

The above equation exhibits a remarkable feature \cite{dk} in the t'Hooft
limit. The leading contributions to both
the l.h.s as well as to the second term of the r.h.s. are $O(\frac{1}{N^2})$.
The connected parts to the first term of the r.h.s. are of order $O(\frac{1}{N^4})$. There is also an  $O(\frac{1}{N^2})$ term built from a disconnected
part which is proportional to the original correlation function if $C^1$
has no intersections. Then, at least in this restricted geometrical setting,
the leading term to the open contour correlator can be expressed in terms
of a closed contour functional.
\section{Further applications of non-commutative Wilson functionals}
There are numerous further instances where Wilson functionals for closed
and open contours play a crucial role. Among them are the construction
of an explicit formula for the Seiberg-Witten map and effective actions
for non-commutative gauge theories. We want to close this short note
with a proposal for a constraint picking out non-commutative $SU(N)$
configurations out of general $U(N)$ configurations \cite{cd}.
It has the same structure as (\ref{9}), only the candidate gauge
invariant operator ${\cal O}$ is replaced by the gauge field $A$
itself
\footnote{Note that in the present review we have chosen path ordering
from right to left.}
\be
\int dx ~ \mbox{tr}\left (U(k,x)\star A(x)\star e^{-i k x}\right )~=~ 0~,
~~~~~\forall k ~.
\label{27}
\ee
The allowed gauge transformations $\Omega (x)$ then have to satisfy
\be
\int dx ~ \mbox{tr}\left (U(k,x)\star \Omega \star d\Omega ^{\dagger} (x)\star e^{-i k x}\right )~=~ 0~,
~~~~~\forall k ~.
\label{28}
\ee
The last condition closes under the composition of two gauge transformations.
\\[8mm]
{\bf Acknowledgments:}\\[2mm]
I thank M. Abou-Zeid and C.-S. Chu for discussions and pleasant collaborations.

\end{document}